\documentstyle[12pt]{article}
\begin{document}
\vspace{0.2cm} \centerline{\Large\bf Density dependence of quark
masses and stability} \centerline{\Large\bf of color-flavor locked
phases }
\vspace{0.2cm} \centerline{ Xiao-Bing Zhang and Xue-Qian Li
} \vspace{0.2cm} \centerline{\small Department of Physics, Nankai
University, Tianjin 300071, China} \vspace{8pt} \vspace{0.2cm}
\begin{minipage}{15cm}
%\centerline
{\noindent   Considering the density dependence of quark masses, we
investigate the color-flavor-locked matter and its stability
relative to the ( unpaired ) strange quark matter. We find that,
when the current mass of strange quark $m_s$ is small, the strange
quark matter remains stable for the moderate baryon densities. When
$m_s$ is large, the gapless phase of the color-flavor-locked matter
is found to be difficult to be stable. A schematic phase diagram of
three-flavor quark matter is presented, in which the
color-flavor-locked phase region is suppressed in comparison with
the previous results.}

\vspace{0.2cm} {PACS number(s): 25.75.Nq, 12.39.Ba, 12.38.-t}
\vspace{0.2cm}
\end{minipage}
\baselineskip 22pt

 \vspace{0.5cm} \noindent {\bf I.
INTRODUCTION}\vspace{0.2cm}

The quark matter with three flavors ( $u$, $d$ and $s$ ) has been
intensively studied for two decades. When the down quark chemical
potential is larger than the strange quark masses, the strange quark
matter ( SQM ) might be energetically favored with respect to
two-flavor quark matter and even nuclear matter so that it should be
the ground state of strongly interacting matter \cite{witt}. Within
the framework of the bag model, Farhi and Jaffe pointed out that SQM
with the strange quark mass $m_s <140$MeV ( and with appropriate bag
constant ) becomes the stable ground state for low baryon densities
\cite{jaff}. Based on this consideration, it is further speculated
that some compact stars are made up not of neutrons but SQM, which
are termed as strange quark stars \cite{ss}. On the other hand, the
study of dense quark matter draws much attention due to the recent
progress in understanding of color superconductivity. At high
densities, the original color and flavor symmetries of three-flavor
QCD, namely $SU(3)_{color}\times SU(3)_{L} \times SU(3)_{R}$, is
suggested to be broken down to a diagonal subgroup
$SU(3)_{color+L+R}$ via the Bardeen-Cooper-Schrieffer ( BCS )
pairing \cite{alf}. The three-flavor quark matter with the
particular symmetry is called the color-flavor locked ( CFL ) matter
and it is different from SQM which is the matter without the BCS
pairing. As another state of strongly interacting matter, the CFL
matter is widely believed to become " absolutely " stable for
sufficiently high densities \cite{alf03}.

Thus there are two candidates for the ground state of three-flavor
quark matter, CFL and SQM, which are stable for high and low
densities respectively. The question is, in the moderate density
region, which one of them is the ground state. In the other words,
one concerns how SQM undergoes a phase transition to CFL with
increase of density. Investigation on these issues is important for
exploring the physics of strange quark stars and/or the interior
structure of compact stars. Ignoring the $u$ and $d$ quark masses,
the CFL free energy takes the form
\begin{eqnarray}
\Omega_{CFL}= - \frac{3\mu^4}{4\pi^2}+ \frac{3m_s^2
\mu^2}{4\pi^2}-\frac{m_s^4-12 m_s^4 \ln\frac{m_s}{2\mu}}{32\pi^2}
-\frac{3\Delta^2 \mu^2}{\pi^2}
 ,\label{pcfl0}
\end{eqnarray}
to the fourth order in $m_s$, where $\mu$ is the quark chemical
potential and $\Delta$ denotes the color superconducting gap. By
comparing Eq.(\ref{pcfl0}) with the free energy of the neutral
unpaired quark matter at the same order, Alford \emph{et. al.}
concluded that CFL is more stable than SQM as long as \cite{alfd}
\begin{equation} {\mu}\geq \frac{m_s^2}{4\Delta}. \label{cflbreak}
\end{equation}
As the necessary condition for the CFL presence, Eq.(\ref{cflbreak})
is valid only for high densities \cite{alfj02}. This inequality can
not fully answer the question raised above because it does not
address the phase structure for moderate densities.

To illustrate this point more clearly, we draw the phase diagram
based on Eq.(\ref{cflbreak}) in the ( $m_s, \mu$ ) plane ( Fig. 1 )
\footnotemark[1] \footnotetext[1] { Until now the actual value of
$\Delta$ and its dependence on $\mu$ and $m_s$ are not been well
known, which are closely linked to the gap equation. In the
literature, $\Delta$ was estimated to be of order tens to $100$MeV.
In Fig. 1  it is simply treated as a given parameter $\Delta \sim
25$MeV for moderate $\mu$, say, $\mu\sim 0.5$GeV \cite{alf04}.}.
When the strange quark mass is small as $m_s <175$MeV, Fig. 1 shows
that SQM is excluded completely from the moderate density region
$\mu=0.3-1$GeV and CFL ( including its gapless phase, see the
following ) dominates all over. However, for small $m_s$, SQM has
been predicted to be the stable ground state \cite{jaff} so that it
should be favorable at least for low densities such as $\mu\sim
0.3$GeV. If assuming that CFL emerges in strange quark stars, this
contradiction becomes more obvious. Starting at very low density and
increasing the matter density by increasing $\mu$, the CFL formation
must be preceded by the appearance of the stable SQM state. From
this point of view, SQM remains stable for relatively low densities;
otherwise, the self-bound quark stars could not exist and then the
CFL formation would be impossible. Therefore, the phase diagram Fig.
1 is problematic especially for relatively low densities.

On the other hand, the so-called gapless CFL phase ( gCFL ) is shown
in Fig. 1, where gCFL separates CFL ( namely the conventional CFL
phase ) and SQM from each other. As suggested by the authors of
Ref.\cite{alf04}, the nonzero value of $m_s$ plays a key role in
triggering the gCFL phase. When $m_s$ is relatively large, thus, it
seems reasonable that gCFL replaces CFL to be more stable in the
moderate density region. However, Casalbuoni \emph{et. al.} have
pointed out recently that gCFL is actually unstable due to the pure
imaginary masses of the gluons in this phase \cite{cas}. It implies
that the gCFL phase region shown in Fig. 1 might be uncorrect. In
addition, the SQM-gCFL transition curve in Fig. 1 was simply
determined by the critical relation of Eq.(\ref{cflbreak}). Since
the validity of Eq.(\ref{cflbreak}) is worthy of doubt for moderate
densities, the gCFL phase region shown in Fig. 1 is problematic
also.

In fact, the implicit assumption for Fig. 1 is that only the current
mass of strange quark $m_s$ was considered in the descriptions for
SQM, CFL and gCFL. According to the low-density QCD, the strange
quark mass not merely originates from the explicit breaking of
chiral symmetry. For low densities where SQM exists as the stable
ground state, there is no reason to neglect the dynamical mass
induced by the spontaneous chiral breaking. Once the dynamical mass
is taken into account, it has been found that the SQM stability
window, e. g. the allowed region of the current mass $m_s$, is
widened \cite{lug,pion}. This motivates us to reexamine the phase
diagram involving SQM and CFL ( gCFL ) in the framework where the
dynamical mass is incorporated. In the present paper, we will
introduce the density dependence of quark mass to investigate the
phenomenological effects of the dynamical mass on the
moderate-density phase diagram. This approach should be closer to
reality and obviously helpful to clarify the problem of Fig. 1. In
Sec.II, we briefly review the mass-density-dependent model
\cite{lug} and consider the free energies of CFL and gCFL when the
density-dependent quark mass is incorporated. In Sec.III, we
investigate the phase transitions from SQM to CFL and/or gCFL and
then present a new phase diagram which is very different from Fig.
1.

\vspace{0.2cm}\noindent {\bf II. THE MODEL }\vspace{0.2cm}

Following Ref.\cite{lug}, the density-dependent quark mass is
given by
\begin{eqnarray} {m}_{D}= {C}/(3\rho),\label{md}
\end{eqnarray}
where $\rho$ denotes the matter density and $C$ is a model
parameter which is constrained by the SQM stability conditions. If
ignoring the current masses of $u$ and $d$ quarks, the masses for
the light and strange quarks in this model are
\begin{eqnarray} {M}_{u}={M}_{d}= {m}_{D};\;\;
{M}_{s}= {m}_{s}+{m}_{D},\label{mq}
\end{eqnarray}
respectively.

The SQM free energy contributed by the Fermi gas reads \cite{lug}
\begin{eqnarray}
{\Omega}(\mu_i,M_i,p^i_F)&=& {{\sum}\atop{i=u,d,s}} \int_0^{p_F^i}
\frac{3}{\pi^2} p^2 (\sqrt{p^2+M_i^2}-\mu_i )dp \nonumber \\
&=& -{{\sum}\atop{i=u,d,s}} \frac{1}{4\pi^2}[{\mu_i}{p_F^i}(
{\mu_i^2}-{5\over2}{M_i^2})+ {3\over2}{M_i^4}\ln
(\frac{\mu_i+p_F^i}{M_i})] .\label{psqm1}
\end{eqnarray}
For each flavor the Fermi momentum ${p_F^i}$ is defined by
$p_F^i=\sqrt{\mu_i^2-M_i^2}$ where $\mu_u=\mu-2\mu_e/3$ and
$\mu_d=\mu_s=\mu+\mu_e/3$ if the electron chemical potential
$\mu_e\neq 0$. On the SQM side, the Fermi momenta of $u$, $d$ and
$s$ quarks are different and are related to the corresponding
densities via $\rho_i={(p_F^i)}^3/{\pi^2}$. Therefore, for SQM,
the electrical neutrality is realized by
\begin{eqnarray}
\frac{2}{3}\rho_u-\frac{1}{3}(\rho_d+\rho_s)=\rho_e=\frac{\mu_e^3}{3\pi^2},
\label{neu}\end{eqnarray} and the baryon density is
\begin{eqnarray} {\rho}&=&\frac{1}{3}(\rho_u+\rho_d+\rho_s).\label{nsqm}
\end{eqnarray}
When the contribution from electrons is included, the total free
energy for SQM becomes
\begin{eqnarray} {\Omega}_{SQM}&=&
{\Omega}(\mu_i,M_i,p_F^i) -\frac{\mu_e^4} {12\pi^2}.\label{psqm}
\end{eqnarray}

With respect to nuclear matter, SQM becomes energetically stable
for low densities as long as its energy per baryon satisfies
\begin{equation} {({\cal E}/\rho)}_{SQM} \le 930\textrm{MeV}, \label{stb}
\end{equation}
at zero pressure, where $930$MeV corresponds to a typical value of
the energy per baryon in nuclei. For our purpose, Eq.(\ref{stb})
needs to be considered seriously to guarantee that not nuclear
matter but SQM undergoes a phase transition to CFL ( if without
this constraint the nuclear-CFL transition \cite{alfd} would be
very likely ). In this work, we fix the parameter $C$ by the
critical condition of Eq.(\ref{stb}) for certainty. For instance,
the value of $C$ is adopted to be $110$ and $70$MeV/fm$^3$ as
$m_s=0$ and $180$MeV respectively ( see Ref.\cite{lug} for details
).

On the other hand, the energy per baryon for two-flavor quark
matter ( 2QM ) is required to satisfy the inequality
\begin{equation} {({\cal E}/\rho)}_{2QM} > 930\textrm{MeV} , \label{1'}\end{equation}
at zero pressure \cite{jaff}. By using Eqs.(\ref{stb}) and
(\ref{1'}), the stability window can be obtained in which SQM
corresponds to the stable ground state at low densities. But
Eq.(\ref{1'}) does not apply to the moderate-density case. Due to
the appearance of strange flavor, SQM is favored over the regular
two-flavor matter as long as
\begin{eqnarray}
\mu_d=\mu+\mu_e/3 \geq M_s.  \label{2s}
\end{eqnarray}
Instead of Eq.(\ref{1'}), thus, Eq.(\ref{2s}) needs to be taken
into account in the following calculation.

Different from the unpaired one, the CFL quark matter is an
insulator in which no electrons are required for the electrical
neutrality \cite{neu}. The Fermi momenta have the common value
\cite{alfd,neu}
\begin{eqnarray}
\nu=2\mu-\sqrt{\mu^2+m_s^2/3},\label{pfcom}
\end{eqnarray}
for all three flavors and $\mu_e$ does not influence the CFL free
energy directly \footnotemark[2] \footnotetext[2] { Assuming that
the SQM-CFL transition is of first order, there is an interface
between the electron-rich SQM and the electron-free CFL. In this
case, the effective value of electron chemical potential is zero on
the CFL side because of the electrostatic potential at the
metal-insulator boundary ( see Refs.\cite{alfd,zhang} for details ).
In the present work, a possibility of the mixed state consisting of
the electrical-opposite SQM and CFL will be ignored. The reason is
that the nonzero $m_D$ provides the additional instanton interaction
so that the electrical-negative phase such as CFL$K^-$ is very
difficult to emerge.}. Thus the CFL free energy contributed by the
Fermi gas is obtained by replacing the variables $\mu_i$ and $p_F^i$
in Eq.(\ref{psqm1}) by $\mu$ and $\nu$ respectively. Together with
the contribution from the CFL pairing ( the last term in the RHS of
Eq.(\ref{pcfl0}) ),
%, namely $-{3\Delta^2 \mu^2}/{\pi^2}$ \cite{alfd},
the total free energy for CFL takes the form
\begin{eqnarray}
\Omega_{CFL}= {\Omega}(\mu,M_i,\nu)-\frac{3\Delta^2 \mu^2}{\pi^2}
,\label{pcfl}
\end{eqnarray}
when the density dependence of quark mass is considered. At high
density, the value of $m_D$ is close to zero so that the difference
between Eqs.(\ref{pcfl0}) and (\ref{pcfl}) becomes negligible. For
the concerned density region, Eq.(\ref{pcfl}) means that not only
$m_s$ but also $m_D$ contribute to the CFL free energy. As a
consequence, the previous results of the SQM-CFL transition e.g.
Eq.(\ref{cflbreak}) are no longer valid at low/moderate densities (
see Sec.III for details ).

Then we turn to consider the gCFL phase in the model where the
density dependence of the quark mass is included. At sufficiently
high densities, it is well known that $m_s^2/(2\mu)$ is regarded as
the chemical potential associated with strangeness, i.e.
$\mu_S=m_s^2/(2\mu)$ \cite{sch}. For the Cooper pairs between the
blue-down ( $bd$ ) and green-strange ( $gs$ ) quasi quarks, the
effective chemical potential for the $gs$ modes is influenced by
$\mu_S$ while that for the $bd$ modes is independent of $\mu_S$. In
this case, the effective chemical potentials $\mu_{gs}^{eff}$ and
$\mu_{bd}^{eff}$ become different and the relative chemical
potential of the paired $bd$ and $gs$ modes becomes \cite{alf04}
\begin{eqnarray}
\delta \mu=\frac{\mu_{bd}^{eff}-\mu_{gs}^{eff}}{2}=
\frac{m_s^2}{2\mu}, \label{dmu0}
\end{eqnarray}
where the contribution from the chemical potential associated with
the color charge has been included \cite{alfj02}. When the variation
$\delta \mu$ is larger than the color superconducting gap, i.e.
\begin{eqnarray}  \frac{m_s^2}{2\mu}\geq \Delta,
\label{gcfl}
\end{eqnarray}
gCFL is more stable than CFL. As a consequence, the free energy
contributed from the gapless phenomenon depends on the comparison
between the variation $\delta \mu$ and the color superconducting gap
$\Delta$ mainly \cite{alfkou,liu}. A natural question is whether the
variation $\delta \mu$ and then the gapless phenomenon are
influenced by the nonzero $m_D$ also. In view of the fact that $m_D$
is the dynamical mass for all three flavors, it can not enter the
strangeness chemical potential $\mu_S$ via a simple replacement $m_s
\rightarrow M_s=m_s+m_D $. Therefore, an extrapolation like $\delta
\mu \rightarrow \delta
\mu'=\frac{M_s^2}{2\mu}=\frac{(m_s+m_D)^2}{2\mu}$ is not feasible in
principle. Thus it is reasonable to assume that $\delta \mu$ is
independent of $m_D$ and Eq.(\ref{dmu0}) holds unchanged when the
density-dependent quark mass is considered. Based on the simple
ansatz, Eq.(\ref{gcfl}) is still valid as the condition for the gCFL
formation although the SQM-gCFL transition needs to be reexamined
seriously.

\vspace{0.2cm}\noindent {\bf III. NUMERICAL RESULTS AND DISCUSSIONS
}\vspace{0.2cm}

As a strong-coupling effect, the nonzero $m_D$ is expected to affect
the phase diagram of three-flavor quark matter for not-very-high
densities. Before going to be specific, let us firstly discuss
\emph{whether or not} the phase transition between SQM and CFL/gCFL
occur in the moderate density region. The answer is not always
positive and it is actually linked to the value of $m_s$, as argued
in the following. Now both SQM and CFL/gCFL are the deconfined
phases, therefore the physics of confinement does not play a role in
determining the SQM-CFL/gCFL transitions. So the negative values of
the free energies obtained in Sec. II are related to the
corresponding pressure directly and then the Gibbs condition for the
pressure equilibrium reads
\begin{eqnarray} P_{CFL/gCFL}-
P_{SQM}={\Omega}_{SQM}-{\Omega}_{CFL/gCFL}=0.
\end{eqnarray}

For $m_s=10$, $50$, $100$ and $150$MeV, we show $\delta
P=P_{CFL}-P_{SQM}$ as a function of $1/\mu$ in Fig. 2. It is found
that $\delta P$ does no longer approach to zero monotonously with
increasing $1/\mu$, i.e. decreasing $\mu$. As shown in Fig. 2, there
exists a rising tendency of $\delta P$ in the vicinity of $\mu\simeq
0.3$GeV. This leads to the fact that no any pressure equilibrium
appears in the moderate density region so that a first-order SQM-CFL
phase transition does not occur. Although the CFL pressure is
relatively large, the absence of the phase transition means that the
CFL matter is impossible to exist at least in our concerned density
region. Therefore, SQM with small $m_s$ still remains as a stable
state for moderate densities while CFL with small $m_s$ is
prohibited unless the density is very large \footnotemark[3]
\footnotetext[3] { At a very high density, the pressures for SQM and
CFL can become close to each other as long as the pairing gap
$\Delta$ is small enough compared with the value of $\mu$. In the
asymptotic sense, the SQM-CFL transition always occur regardless of
whether $m_s$ is small. But this is not the case being concerned in
the present work. }.

The above physical picture holds valid until $m_s$ is large. For
larger $m_s$, more pressure is paid to maintain the common Fermi
momentum so that the pressure of CFL decreases. As long as $m_s$ is
large enough, the SQM-CFL transition in moderate density region
becomes possible to occur. Our numerical calculation shows that, as
$m_s$ is about $150$MeV, the pressure equilibrium comes to appear in
the vicinity of $\mu\simeq 0.3$GeV ( see Fig. 2 also ). Noticing
that such pressure equilibrium behaves like " crossover " of CFL and
SQM, $m_s \simeq150$MeV is regarded as the minimal value allowed for
the CFL existence at moderate density. Once $m_s$ is larger than
$150$MeV, the transition from SQM to CFL occurs in moderate density
region. As a typical example, the result of $\delta
P=P_{CFL}-P_{SQM}$ for $m_s=200$MeV is given by the solid line in
Fig. 3. As shown in Fig. 3, the critical chemical potential $\mu_c$
for the SQM-CFL transition is about $0.4$GeV. This means that, for
$m_s=200$MeV, SQM remains stable in the region of $\mu < \mu_c$ and
CFL emerges as $\mu \geq\mu_c$.

Then, we turn to consider the possibility of the SQM-gCFL phase
transition. When $m_s$ is small such as $m_s<150$MeV, the CFL matter
is not stable for moderate densities so that its gapless phase does
not exist at all. Even if $m_s$ is large, we emphasize that the
presence of gCFL is still difficult because of the absence of the
SQM-gCFL transition. For illustrative purpose, let us examine the
gCFL phase in the case of $m_s=200$MeV. In Fig. 3, the critical
chemical potential $\mu_g$ for the gCFL formation ( i. e. the
gCFL-CFL transition ) is shown to be about $0.67$GeV, which is
obviously larger than the critical value $\mu_c$ for the SQM-CFL
transition. Since the gapless phenomenon is determined by
$m_s^2/(2\mu)$ mainly, the difference between the gCFL and CFL
pressures rises with increasing $1/\mu$. Therefore, the value of
$P_{gCFL}-P_{SQM}$ does not approach to zero even if $\mu$ is close
to $\mu_c$, as shown by the dashed line in Fig. 3. As a result,
there is not the pressure equilibrium between gCFL and SQM so that
the phase transition to gCFL actually does not occur in the quark
star environment. Although gCFL is more energetically favorable than
CFL, thus, it is impossible to emerge as the stable state for
$m_s=200$MeV.

The above argument is similar as that involving the CFL unstability
in the case of $m_s$ is small ( Fig. 2 ) and it holds valid for the
most values of $m_s$. For example, if $m_s$ is larger than $200$MeV,
more pressure is gained from the gapless phenomenon and the raising
tendency of $P_{gCFL}-P_{SQM}$ becomes more obvious as $\mu$ is
small. So the SQM-gCFL transition and then the presence of gCFL are
prohibited. The exception happens only if the critical points
$\mu_g$ and $\mu_c$ are close to each other. In that case, the
difference between the gCFL and CFL pressures becomes small, so that
the SQM-gCFL transition occurs also and the corresponding chemical
potential approaches to $\mu_c$. This implies that gCFL emerges as
the stable ( exactly, metastable ) state for a very narrow region of
quark chemical potential. Our numerical calculation shows that such
gCFL phase is impossible unless $(m_s,\mu)$ is limit in the vicinity
of ( $183$MeV, $0.37$GeV ), which corresponds to the intersection of
the SQM-CFL and CFL-gCFL transitions ( see Fig. 4 also ).

Based on the above arguments, a schematic phase diagram of the CFL
quark matter is given for the moderate density region in Fig. 4.
There are three kinds of different structures in the phase diagram
according to the value of $m_s$ :

(i) As $m_s$ is small such as $m_s <150$MeV , the effect of $m_D$
prohibits the CFL formation for not-very-high densities. In this
case, it is not CFL but SQM to be the stable state in the whole
moderate density region, as shown in Fig. 4. This conclusion is very
different from that obtained by Fig. 1, but agrees with the original
prediction that CFL with zero ( or small ) $m_s$ becomes possible
only when the density is high enough \cite{alf}.

(ii) When $m_s$ is larger than $150$MeV, a first-order transition
from SQM to CFL takes place. Due to the density-dependent quark mass
$m_D$, the SQM-CFL transition curve shown in Fig. 4 is deviated from
the relation Eq.(\ref{cflbreak}) obviously. With respect to the
result of Fig. 1, we find that the CFL phase region is strongly
suppressed for moderate densities. More importantly, the effect of
$m_D$ prohibits the presence of the so-called gCFL phase also. For
the most values of $m_s$, we find that SQM does not undergo a first
order phase transition to gCFL and thus gCFL can not emerge as the
stable state in the strange quark-star environment. This point is
qualitatively different from the result of Fig. 1, in which gCFL was
shown to be dominant as long as $m_s$ is relatively large. With
respect to the result of Fig. 1, we find that the presence of the
stable gCFL phase is reduced to almost nil.

(iii) As $m_s$ is very large, the existence of three-flavor quark
matter including SQM becomes difficult. By using Eq.(\ref{2s}) we
give the boundary curve of SQM ( namely the 2QM-SQM transition curve
) in Fig. 4, which shows that 2QM seems irrelevant to the presence
of CFL. Once the color superconductivity is incorporated into 2QM,
however, the possibilities of the 2QM-CFL/gCFL transitions could not
be ruled out simply \cite{g2sc}. In that case, the dashed line shown
in Fig. 4 needs to be modified, which is beyond the scope of the
present work.

In summary, we extend the description of color-flavor-locked matter
from high-density case to the moderate density region where the
density dependence of quark masses can not be ignored. Starting at
low density and raising the matter density, the physical picture
that SQM remains as the stable state at first and then undergoes a
first-order phase transition to CFL is reexamined in details. We
predict a very different phase diagram of three-flavor quark matter,
in which both the CFL and gCFL phase regions are suppressed for
moderate densities. The present phase diagram is helpful to better
understand the ground states of strongly interacting matter
especially in the environment of strange quark stars. Of course,
there are some uncertainties of the color superconducting gap used
in this work. When the value of the gap is chosen in other ways, we
can give the similar phase diagram as Fig. 4. For instance, if the
gap is large such as $\Delta\sim 80$MeV \cite{kap} we find that the
minimum value of $m_s$ allowed for the CFL existence increases so
that the CFL phase region for moderate densities might be further
suppressed. Even if the density dependence of the gap is included,
the change of $\Delta$ in the finite region of $\mu=0.3-1$GeV is not
too drastic and the conclusion obtained from Fig. 4 is still
qualitatively correct. In the further work one should construct the
dynamical quark mass within a more realistic framework such as that
beyond the bag model as well as take the two-flavor color
superconducting phases into account. Some of the problems are being
investigated.

\vspace{0.5cm} \noindent {\bf Acknowledgements} \vspace{0.5cm}

This work was supported by National Natural Science Foundation of
China ( NSFC ) under Contract No.10405012.

\newpage
\begin{figure}
\caption{Schematic phase diagram of the CFL quark matter in the (
$m_s$,$\mu$ ) plane, where the solid line is obtained from
Eq.(\ref{cflbreak}) and the dashed line corresponds to the phase
transition from the conventional CFL to gCFL phases ( see
Eq.(\ref{gcfl}) in the following ).}\label{1}
\end{figure}

\begin{figure}
\caption{The CFL pressure vs. the SQM pressure. The solid lines
from top to bottom are the relative pressures for $m_s=10$, $50$,
$100$ and $150$MeV respectively.}\label{2}
\end{figure}

\begin{figure}
\caption{The CFL/gCFL pressures vs. the SQM pressure for
$m_s=200$MeV, where the solid and dashed lines are the relative
pressures for CFL and gCFL respectively.}\label{3}
\end{figure}

\begin{figure}
 \caption{ Similar as Fig. 1 but the SQM-CFL transition is
considered in the case of including effects of $m_D$. The solid line
is the boundary of color-flavor-locked matter while the dashed line
is the boundary of three-flavor quark matter. The square marks the
possible region for the gCFL existence as the metastable
phase.}\label{4}
\end{figure}

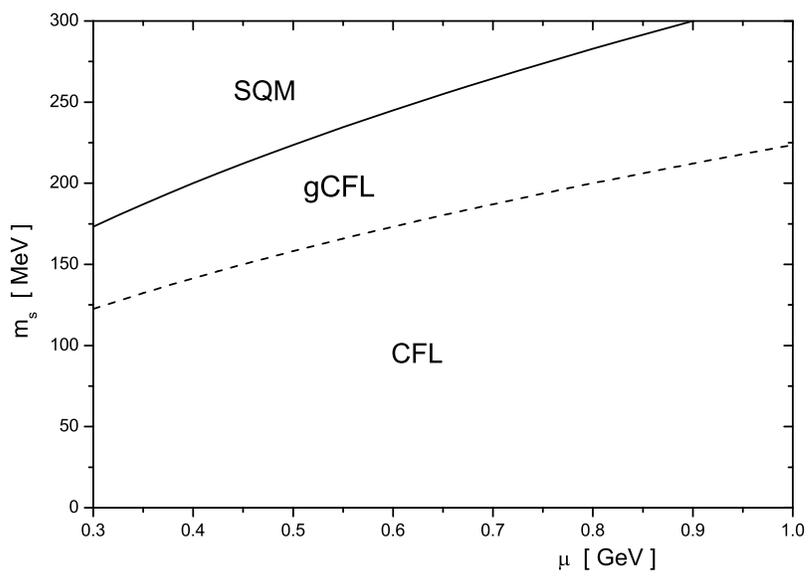
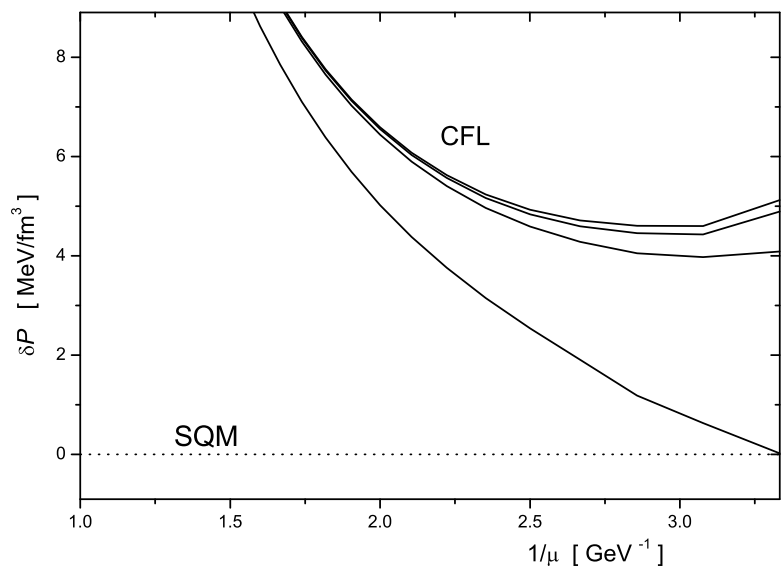
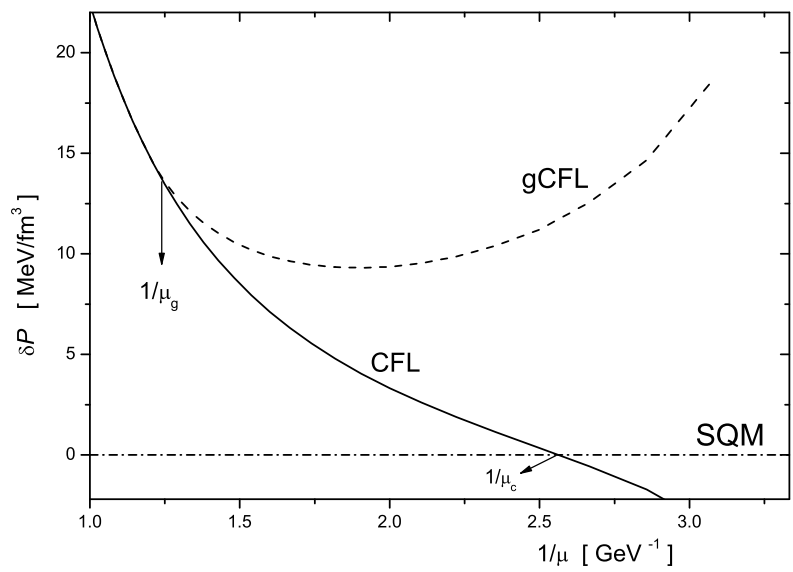
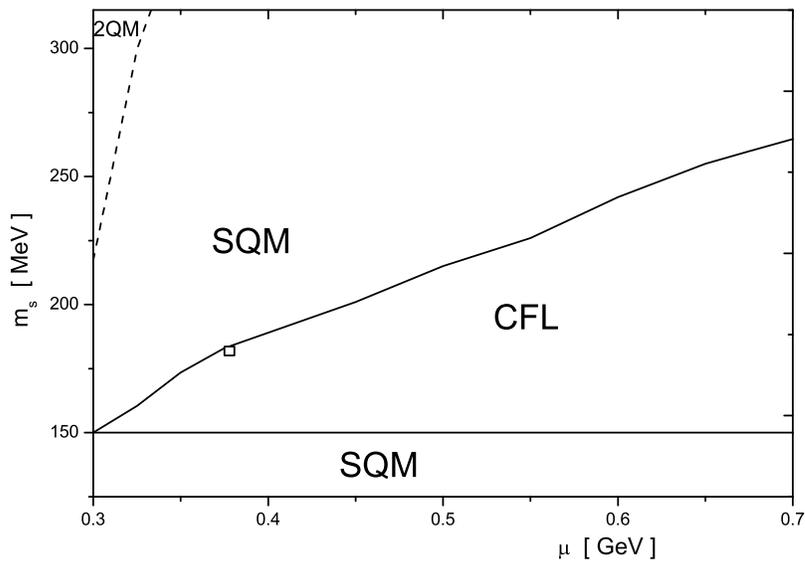
\end{document}